# A first-principles model of early evolution: Emergence of gene families, species and preferred protein folds

Konstantin B. Zeldovich[1], Peiqiu Chen[1,2], Boris E. Shakhnovich[3], Eugene I. Shakhnovich[1]

[1]Department of Chemistry and Chemical Biology,  [2]Department of Physics, and [3]Molecular and Cellular Biology, Harvard University, 12 Oxford Street, Cambridge, MA 02138

Abstract

In this work we develop a microscopic physical model of early evolution, where phenotype - organism life expectancy - is directly related to genotype – the stability of its proteins in their native conformations which can be determined exactly in the model. Simulating the model on a computer, we consistently observe the ''Big Bang'' scenario whereby exponential population growth ensues as soon as favorable sequence-structure combinations (precursors of stable proteins) are discovered. Upon that, random diversity of the structural space abruptly collapses into a small set of preferred proteins. We observe that protein folds remain stable and abundant in the population at time scales much greater than mutation or organism lifetime, and the distribution of the lifetimes of dominant folds in a population approximately follows a power law. The separation of evolutionary time scales between discovery of new folds and generation of new sequences gives rise to emergence of protein families and superfamilies whose sizes are power-law distributed, closely matching the same distributions for real proteins.  On the population level we observe emergence of species – subpopulations which carry similar genomes. Further we present a simple theory that relates stability of evolving proteins to the sizes of emerging genomes. Together, these results provide a microscopic first principles picture of how first gene families developed in the course of early evolution.

**Synopsis**

Here we address the question of how Darwinian evolution of organisms determines molecular evolution of their proteins and genomes. We developed a microscopic ab inito model of early biological evolution, where the fitness (essentially lifetime) of an organism is explicitly related to the evolving sequences of its proteins. The main assumption of the model is that the death rate of an organism is determined by the stability of the least stable of their proteins. A lattice model is used to calculate stability of all proteins in a genome from their amino acid sequence. The simulation of the model starts from 100 identical organisms each carrying the same random gene and proceeds via random mutations, gene duplication, organism births via replication and organism deaths. We find that exponential population growth is possible only after the discovery of a very small number of specific advantageous protein structures. The number of genes in the evolving organisms depends on the mutation rate, demonstrating the intricate relationship between the genome sizes and protein stability requirements. Further, the model explains the observed power-law distributions of protein family and superfamily sizes, as well as the scale-free character of protein structural similarity graphs. Together these results and their analysis suggest a plausible comprehensive scenario of emergence of protein universe in early biological evolution.

**Introduction**

Molecular biology has collected a wealth of quantitative data on protein sequences and structures, revealing complex patterns of the protein universe, such as markedly uneven usage of protein folds and near-scale-free character of protein similarity networks [1-5].On a much higher level of biological hierarchy, ecology, evolution theory, and population genetics established a framework for studying speciation, population dynamics and other large-scale biological phenomena [6-8]. While it is widely accepted that gene families and the Protein Universe emerged during the course of molecular evolution through selection [9-11], there is a substantial gap in our conceptual and mechanistic understanding of how molecular evolution occurred or what are the determinants of selection. Indeed, evolution, as we understand it, proceeds at the level of organisms and populations but not at the level of genomes. Evolutionary selection at the molecular level occurs due to a relation between genotype and phenotype, although a detailed understanding of this relation and its consequences for molecular evolution remains elusive.

A number of phenomenological models (e.g. Eigen's quasispecies model) were developed where fitness of an organism was related to the sequence of its genome. [12] [13,14]. A standard definition of fitness in phenomenological models is the growth rate of a population which is higher for the more fit species. However, the relationship between genotype and phenotype in quasispecies and similar population genetics models is purely phenomenological. For example, in single-fitness peak models, one specific genotype is postulated to be most fit while deviations from it confer selective disadvantage. Despite providing several important insights, these types of approaches lack a fundamental microscopic connection between fitness and easily justifiable, on biological grounds, and measurable quantities (e.g. structure/stability, function or regulation) of proteins. Therefore, such models cannot accurately describe molecular evolution of proteins.

On the other hand, a number of models were proposed that focus on emergence and evolution of sequences of model proteins and RNA under direct pressure on their molecular properties such as stability [11,15-18], folding kinetics [19,20] and mutational robustness [21]. Schuster and Stadler [22] first studied evolution of biological macromolecules – RNA - in the context of population dynamics. Later, in a series of papers [10] Taverna and Goldstein used Eigen model of reaction flow to grow populations of proteins modeled as two-dimensional 25-mers. These authors showed that when requirement to exceed certain stability threshold is imposed the resulting distribution of structures in the evolved population appears highly skewed towards more designable [23] structures and more robust (i.e. less susceptible to mutations) proteins [24].

One of the most surprising features of the Protein Universe, is an uneven and broad distribution of proteins over folds, families and superfamilies. While this fact had been noted by many researchers long ago [1,4,25,26], the quantitative descriptions of these distributions began to emerge only recently. Huynen and van Nimwegen found that sizes of paralogous gene families follow a power-law distribution [3]. Gerstein and coworkers [5] observed power-law distribution of frequencies of several other properties of gene families as defined in the SCOP database [27]. Dokholyan et al [2] studied a network of structural similarities between protein domains (called the Protein Domain Universe Graph, or the PDUG) and found that distribution of connectivities within the PDUG follows a power law (within a limited range of connectivity variance) making it a finite size counterpart of a scale-free network. This is in striking variance with an expectation from random distribution of folds which would result in an (approximately) Gaussian distribution of connectivities of the PDUG .

Ubiquitous nature of power law dependencies of many characteristics of gene families and Protein Universe may suggest their possible common origin from the fundamental evolutionary dynamics and/or physics of proteins. Huynen and van Nimwegen [3], Gerstein and coworkers [5] and Koonin and coworkers [9] [28,29] proposed dynamical models (the version proposed in[28] is called BDIM) based on gene duplication as a main mechanism of creation of novel types. Such models, while providing power-law distribution of family sizes in some asymptotic cases, are sometimes based on assumptions that call into question their generality. In particular, as pointed out by Koonin and coworkers, in order for gene duplication dynamic models to provide non-trivial power law distributions of paralogous family sizes, one has to assume that the probability of gene duplication *per gene* depends, in a certain regular way, on the size of already existing gene family. Further, even under this assumption the power-law distribution in the BDIM model arises only asymptotically in a steady state of evolutionary dynamics [29]. In contrast, the duplication and divergence phenomenological model of Dokholyan et al [2] did not use such dramatic assumptions. However, this model is limited to explanation of scale-free nature of the PDUG and it does not provide any insight as to nature of power-law distribution of gene family sizes. In protein sequence space, a similar approach has been employed by Qian et al [5]. However, models like the ones proposed in [2,3,5,28] and other works are purely phenomenological in nature whereby proteins are presented as abstract nodes and where sequence-structure relationships are not considered.

Here, we present a microscopic physics-based model of early biological evolution (Figure 1) with realistic generic population dynamics scenario where fitness (i.e. life expectancy) of an organism is related to a simple necessary requirement of functionality of its proteins - their ability to be in native conformations. Since the latter can be estimated exactly in our model from sequences of evolving genomes, the proposed model provides a rigorous, microscopic connection between molecular evolution and population dynamics. We demonstrate that the model indeed bridges multiple evolutionary time scales, thus providing an insight into how selection of a best-fit phenotype results in molecular selection of proteins and formation of stable, long-lasting protein folds and superfamilies. Furthermore, the coupling of molecular and organismal/populational scales results in the emergence of species – subpopulations of evolved organisms whose genomes are similar within their groups and dissimilar between groups. The resulting Protein Universe features power-law distribution for gene family and superfamily sizes closely matching real ones. The proposed model can be viewed as a first step towards a microscopic, first principles description of emergence and evolution of Protein Universe.

**Figure 1.** Schematic representation of the genome and population dynamics in the model. Individual genes undergo mutations and duplications. Organisms as a whole can replicate, passing their genomes to the progeny, or die, effectively discarding the genome.

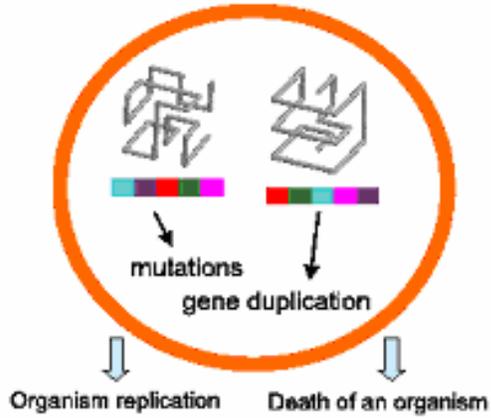

**Figure 1.** *Schematic representation of the genome population dynamics in the model. Individual genes undergo mutations and duplications. Organisms as a whole can replicate, passing their genomes to the progeny, or die, effectively discarding the genome.*

**Results and Discussion.**

**Population dynamics, fold discovery, and punctuated evolution.**

Our evolution dynamics runs start from initial population of 100 organisms each having the same one primordial gene in their genomes. Initial gene sequence is random. Runs proceed according to evolutionary dynamics rules as described in Model and Methods section (see also Figure 1). The life expectancy of an organism is directly related to stability of its proteins as explained in Methods section; briefly, the death rate $d$ is inversely related to protein stability,

$$d = d_0 \left(1 - \min_i P_{nat}^{(i)}\right) \qquad (1)$$

This equation expresses a postulate that all genes of early organisms were essential at the given time; no *a priori* assumptions about the number of these genes are made.

We found that out of 50 simulation runs starting with different starting sequences, 27 runs successfully resulted in a steady exponential growth of the population, whereas in 23 runs the population has quickly gone extinct. A typical behavior of the population growth and protein structure dynamics in a successful evolution run is shown in Figure 2. After a period of "hesitation" lasting for about 100 time steps, a steady exponential growth of the population sets in (Figure 2b). In Figure 2c, we present the mean native state probability $<P_{nat}>$ of all proteins present in the population at a given time. Due to mutations and selection, $<P_{nat}>$ steadily increases with time, and dramatically exceeds the mean $P_{nat}$ for random sequences, $<P_{nat}^{rand}>=0.23$. In contrast to earlier models [10] the selection pressure is applied to whole organisms rather than to individual protein molecules. The genotype-phenotype feedback, which we model by eq. (1) (see also Methods), transfers the pressure from organisms to individual proteins to gene sequences. Figures 2(b,c) show that our selection mechanism works and results in the discovery of stable proteins due to evolutionary pressure.

Using our model, we can follow each structure in the population. In Figure 2a color hue encodes the number of genes in the population corresponding to each of the 103346 lattice structures (ordinate) as a function of time (abscissa). Structures marked in green are the most abundant in population at a given time, while black background corresponds to structures not found in any of the evolving organisms. The most important feature of this plot is the appearance of specific structures that correspond to highly abundant proteins comprising a significant fraction of the gene repertoire of the population. In what follows we will call them Dominant Protein Structures (DPS). Such proteins visually appear as bright lines on Figure 2a. What is the genesis of DPS and how is their appearance related to population growth or decay?

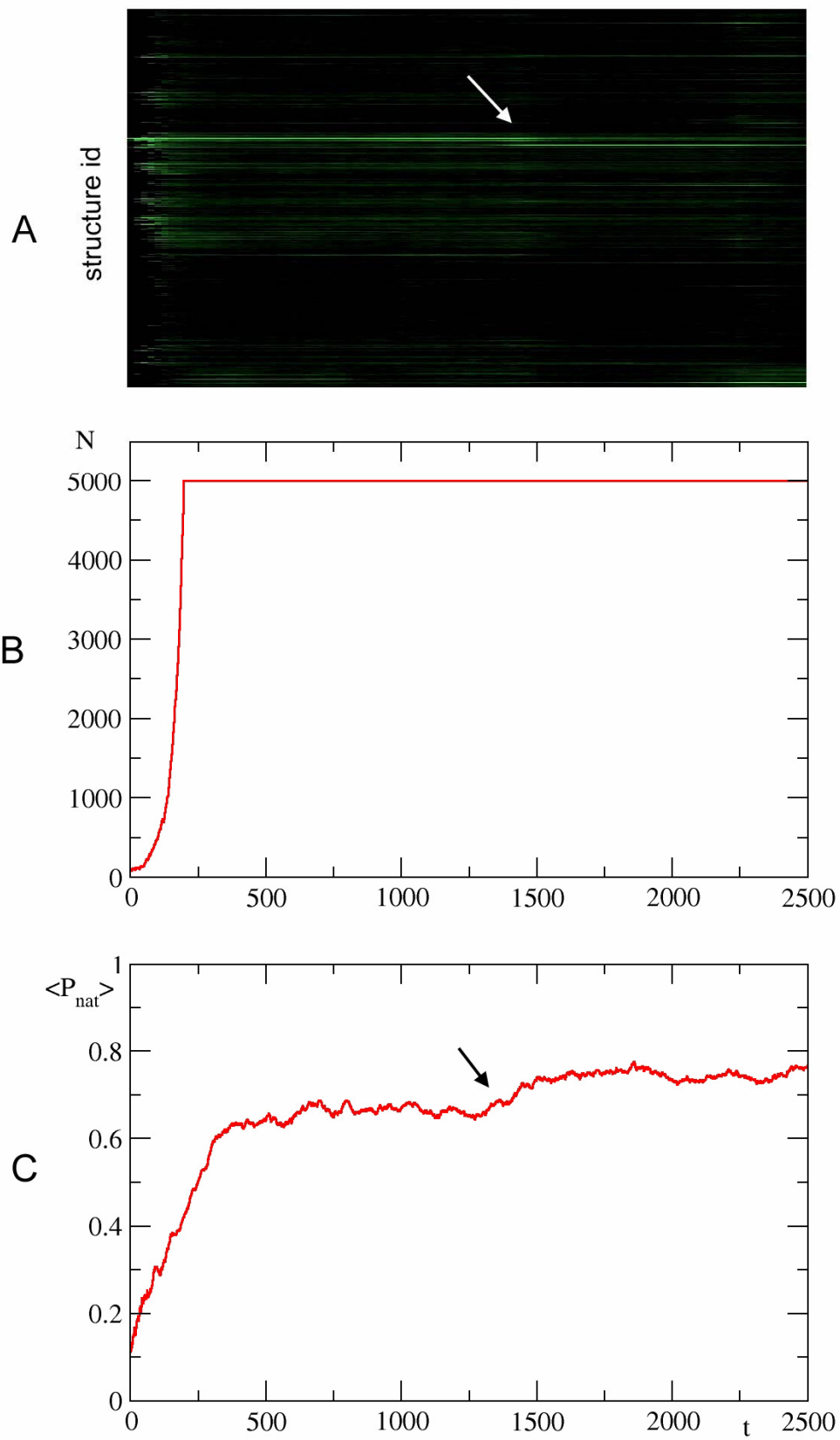

Figure 2.

*Time progression of an evolution run. **(A)**. Structural repertoire of an exponentially growing population as a function of time (abscissa). Ordinate represents the number of the structure out of the 103346 possibles, and abundance of a structure at a given time is encoded by color. Bright green corresponds to abundant structures, and black to rare or nonexistent ones. Arrows point to the discoveries of dominant protein structures (DPS, "bright lines" in the structure repertoire). **(B)** Population as a function of time. Exponential growth sets in as soon as stable dominant protein structures have been found. **(c)** Mean native state probability $<P_{nat}>$, an equivalent of mean population fitness as a function of time.*

To answer this question, let us track the development of the population of structures in time by comparing the structure repertoire, the population size and $<P_{nat}>$ plots. At $t=0$ the proteome consists of a single sequence-structure combination (a single line on the structural repertoire plot) which corresponds to all individuals in the initial population having that structure in the genome. Over time, random mutations diverge sequences in each organism such that the dominance of a single structure is lost. This can be seen as a smeared line on the structural repertoire plot, as shown in Figure 2a, $t<100$. However, at a certain point, very favorable sequence-structure combinations are discovered. They represent DPS whose incorporation into the genome leads to an abrupt increase of $<P_{nat}>$ and explosive exponential growth of the population through increase in fitness. Shortly after the discovery of that DPS, the diversity of the structural space abruptly collapsed, as most of the organisms converge towards the newly discovered DPS. Such a dramatic event – discovery of a limited number of dominant proteins and ensuing exponential growth of the population - can be called the ''Biological Big Bang'', following a loose analogy with astrophysics. As seen on Figure 2a, the emerged dominant folds are very persistent in time. Nevertheless, fold discovery can occur at later stages of evolution. For example, in this particular simulation, at $t\sim1300$, new folds were discovered (white arrow in Figure 2a), they become new DPS and the initial DPS are completely replaced by the new folds by $t\sim1600$. This switchover, accompanied by a marked increase of $<P_{nat}>$, is a clear manifestation of punctuated discoveries of new folds coupled with selection at the organismal level.

Even though the number of organisms increases exponentially, the number of genes in each genome increases very slowly (and stabilizes after the discovery of DPS (Supplementary Figure 1, red curve). Indeed, large genomes are not very advantageous in our model, as mutations occur in all of the genes whereas the death rate is controlled by the gene with the lowest $P_{nat}$. Thus, it is only this gene that bears the brunt of selective pressure. Therefore, the rest of the genome accumulates mutations and is more prone to deleterious mutations. Unless all of the genes are very carefully selected (or formation of pseudogenes is allowed), a larger number of genes means that there is a substantial probability that a point mutation will result in a sequence-structure combination with a very low $P_{nat}$, immediately killing the organism. The observed slow increase of the size of the genome reflects the subtle balance between the selection pressure and gene duplication and is analyzed in more detail below. Remarkably, the average number of genes in the surviving organisms decreases with increasing mutation rates (Supplementary Figure 7). Indeed, if every protein is essential, then the probability of organism death due to a deleterious mutation is lower in organisms with shorter genomes. This result allows direct experimental verification, and clearly sets the current model apart from the previous sequence evolution simulations [10], which focused on the properties of individual proteins.

   Supplementary Figure 2 shows the structural repertoire and population size of an unsuccessful simulation run, where the population quickly became extinct. This simulation did not result in a discovery of a stable fold, and the structural space was evenly filled till the extinction of the population. We found (data not shown) that the choice of starting sequence does not have any significance in determining whether a particular simulation run will result in exponential growth or extinction. Furthermore, in the case of most unsuccessful evolution runs, the genome size rapidly increases with time (Supplementary Figure 1, blue curve), decreasing the average evolutionary pressure per gene and making the discovery of DPS less likely.

   Based on these observations, we conjecture that biological evolution, exponential population growth, and existence of stable genomes are possible only after the discovery of a narrow set of specific protein structures.

## Emergence of Families and Superfamilies

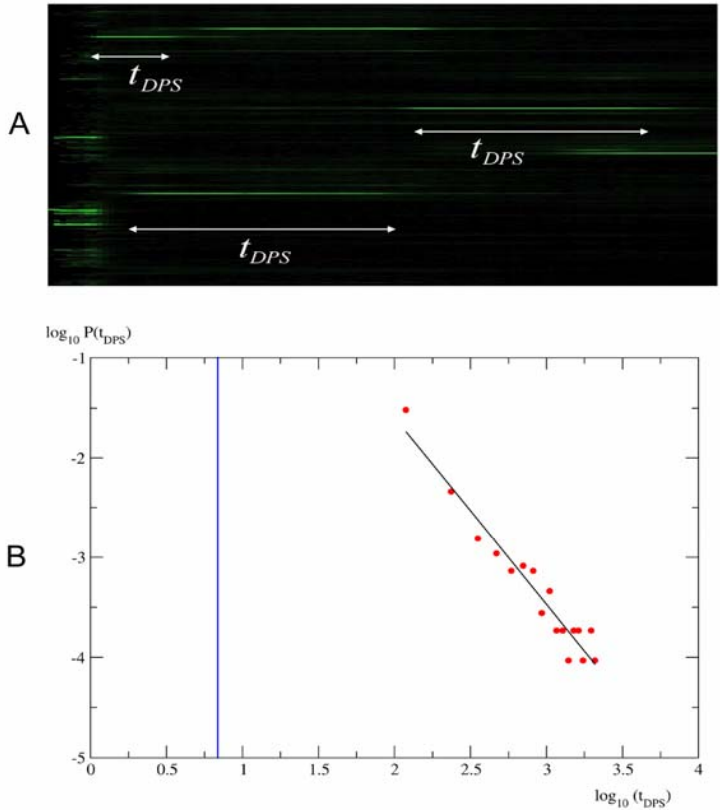

**Figure 3.** *Distribution of life times of DPS. **(A)** Lifetimes are defined as a span between emergence of a DPS when it takes over at least 20% of gene population (seen as bright line here) till its extinction as a DPS when it no longer dominates the population. **(B)** The lifetime distribution of DPS approximately follows a power law with exponent -1.87. DPS folds persist over many generations and eventually give rise to protein superfamilies. The blue line indicates the mean life time of an organism.*

To quantify the persistence of the DPS during evolution, we calculated the distribution of DPS lifetimes, defined as the time span during which a structure comprises more than 20% of the genes present in the most populated structure – i.e. time between emergence of a DPS and its extinction in the population (see Figure 3a).

We consider only DPSs that already completed their ''lifecycle'' i.e. the DPS that emerged and went extinct over the time of an evolutionary simulation. It is clear from Figure 3b that the life-time of a DPS is much greater than that of an organism, or the average time between successive mutations. Moreover, the distribution of DPS lifetimes clearly follows a power-law-like distribution. The long non-exponential tail of the distribution demonstrates that some protein folds are extremely resistant to mutations and may persist over thousands of generations. Over such a long time, diverse protein (super)families are formed around the DPS folds. This is illustrated on Figure 4a which shows the distribution of sizes of evolved families and superfamilies of proteins. To avoid confusion we note that families and superfamilies here are defined not necessarily as sets of paralogous sequences but in the same way as they are defined in SCOP [27]: protein families are defined as sets of all (not necessarily belonging to the same organism) homologous sequences that fold into a given domain structure and superfamilies are defined as all monophyletic sets of sequences whose homology may not be detectable by sequence comparison methods but which nevertheless fold into structurally similar domains. The statistics of protein families is dominated by orthologous genes, in contrast to paralogous families studied in [5,9]. As shown in Figure 4a, both family and superfamily size distributions of evolved proteins follow almost perfect power laws with power law exponent being greater for superfamilies (-2.92) than that for families (-1.77).

In order to compare this result with real proteins we plotted the distribution of family and superfamily sizes of real proteins. As a measure of family sizes here we estimated the number of homologous sequences that fold into a given domain (see Methods) and as a proxy for superfamily size we estimated the number of functions performed by each domain. Clearly the distributions in Figure 4b follow power-law statistics, and as in model, the exponent for the superfamily distribution (-2.2) is greater than that for families (-1.6). Quantitatively, the slopes of the model and real distributions are similar.

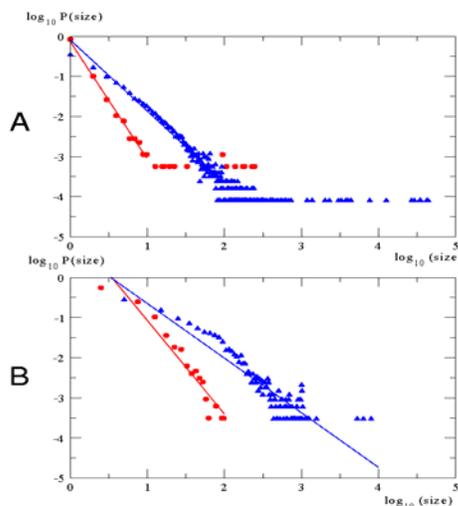

**Figure 4.** *Distribution of family and superfamily sizes **(A)** model evolution. The blue triangles represent the number of sequences folding into the same structure (gene family); the blue solid line approximates a power law with exponent -1.77. The red circles represent the distribution of the number of nonhomologous (Hamming distance greater than 56%) sequences folding into the same structure (superfamilies). The red solid line is a power law with exponent -2.92. **(B)** Orthologous gene family and superfamily sizes*

*in real proteins. The red circles are the number of different functions performed by each domain as defined by InterPro( Bin size =2, Pearson R= .97 of fit with slope = -2.2) and theblue squares are the number of non-redundant sequences folding into each domain. (Bin size = 10, Pearson R=.92 of fit with slope = -1.5).*

**Genome sizes of model organisms**

As mentioned above, the genomes of model organisms from exponentially growing populations are rather short (about 3 genes), in contrast to the extinct populations, where uncontrolled gene duplication is observed (Supplementary Figure 1). To better understand this phenomenon, one should consider the distribution of protein stabilities $P_{nat}$ before and after a round of mutations. Suppose each genome has $N$ genes, and the fitness (inversely related to the probability of death of an organism) of the genome is defined by $f = \min\{P_{nat}^{(1)}, \ldots, P_{nat}^{(N)}\}$. At each time step, each gene in the genome has an equal probability of mutation. For simplicity, we assume that the distribution of stability $P_{nat}$ of a lattice protein after a point mutation (i) does not depend on the stability before the mutation, and (ii) is uniformly distributed between 0 and 1. Such a crude approximation works surprisingly well for lattice proteins (see Supplementary Figure 3), and allows for an analytic calculation of the average genome fitness $f'$ after a round of point mutations (see Methods). The average fitness after a point mutation depends on the number of genes $N$ and original fitness $f$, and since larger genomes accumulate more mutations, they are more prone to a decrease in fitness after the mutation. In particular, if the fitness of an original genome was $f$, then after one round of point mutations the average new fitness $f'$ will be no less than $f$ only if the genome is sufficiently short, namely if

$$N < \frac{3f + \sqrt{17f^2 - 16f + 8}}{2f} \qquad (2)$$

In other words, on average, organisms with more than $N$ genes will decrease their fitness after a point mutation and will be eventually washed out from the population. Thus, eq. (2) establishes an upper boundary on the number of genes per organism at a given level of stability $f = \min\{P_{nat}^{(i)}\}$ in the weakest link model of evolution with lattice proteins. In Figure 5, we plotted the predicted boundary from eq. (2) and the results of 50 simulation runs, where we show the scatter between the average number of genes per organism $N$ in a population and average stability $<P_{nat}>$ of proteins in a population at every time step during each of the runs. As predicted by eq. (2), only organisms with sufficiently short genomes survive at a given level of protein stability; the higher is the stability, the lower is the maximum possible number of genes per organism. Indeed, in a genome consisting of very stable proteins most of the mutations are deleterious and confer a lethal phenotype in our evolutionary model. In this particular model, no more than 3 genes can be present in a genome at very high values of $P_{nat}$. It should be noted that this consideration applies only to the equilibrium size of the genome at large evolutionary times, and does not describe the entire course of its evolution in time (Supplementary Figure 1).

A more realistic distribution of changes of $P_{nat}$ upon mutation and a more detailed consideration of effect of mutations on the fate of the organisms result in realistic estimates of genome sizes for real organisms (KBZ,PC,EIS, manuscript in preparation).

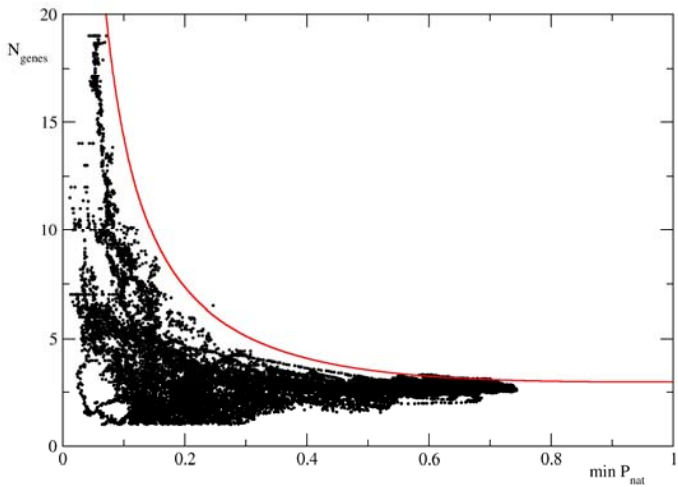

**Figure 5.** *Analytic prediction for the maximum number of genes in an organism as function of the mean protein stability $P_{nat}$ (f in equation (3)) in the weak link lattice model (red curve) and the results of simulations (black dots). The data from 50 simulation runs, both exponentially growing and extinct, have been combined.*

**Emergence of clonal lines or species**

The exact nature of the model gives us direct access to the genomes of all evolved organisms, and an interesting question is whether all evolved genomes are similar (monoclonal, or single-species population), and if not, can they be clustered into distinct clonal lines or species. It turns out that the number of DPS in the evolved population is a very good indicator of species formation. In many cases, there is only one DPS in the evolved population. Then, the genomes of all organisms are similar, and the population is monoclonal. A more interesting case is presented in Figure 6(a), where two different DPS, corresponding to structures 10107 ("A") and 15550 ("B") (in our arbitrary numbering) have evolved.

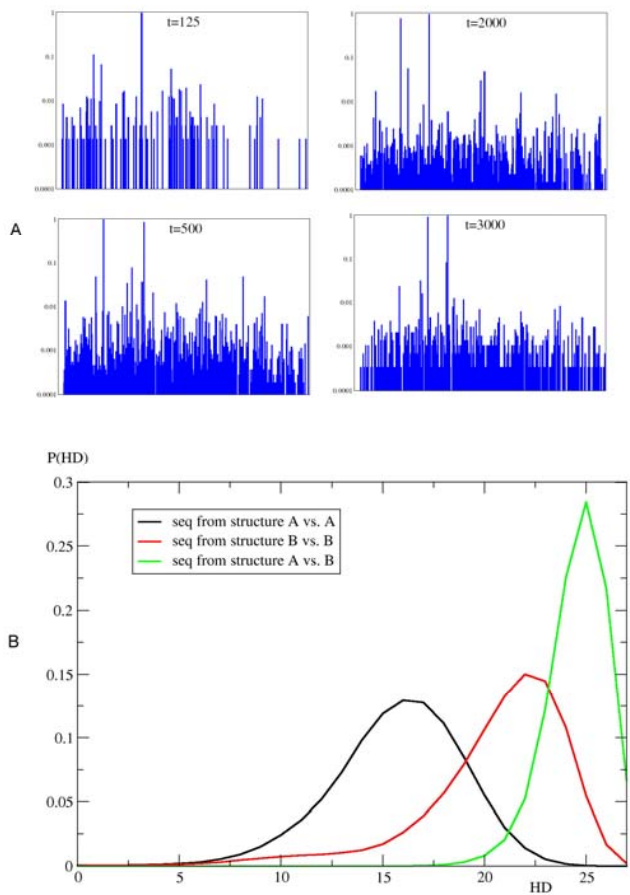

**Figure 6.**

*Emergence of species. (A) Structural repertoire of an evolution run developing two dominant protein structures. The height of the bars represents the number of sequences folding into a given structure; the structure numbers are arbitrary.*

*(B) Histograms of pairwise Hamming distances between sequences corresponding to the two DPS (black and red curves) demonstrate sequence similarity within the structure's superfamily. The histogram of Hamming distances between the sequences folding into one DPS and the sequences folding into another DPS (green) shows a lack of sequence similarity. As each organism bears only one of the two DPS, one can say that this evolution run resulted in the formation of two different strains, or species, of organisms.*

Are these structures randomly distributed between organisms, or are there are groups of organisms preferentially using structure A, but not B, and vice versa? In the latter case, one could argue that two clonal lines, or species, have evolved, as each of the groups will have its own and distinct set of protein structures, and, correspondingly, sequences. It turns out that in 1536 organisms in the population, at least one gene encodes for the structure A, in 2767 organisms, at least one gene encodes for B, but there are *no* organisms that include both A and B in their genomes. 697 organisms have neither A nor B in their genomes. Therefore, organisms having the DPS of fold A in their genomes are very distinct from the organisms with the fold B. This difference is further illustrated in Figure 6(b), where we plotted the histograms of pairwise Hamming distances between the sequences encoding for the structures A and B. The black curve represents the distribution of all pairwise Hamming distances between the sequences encoding for the structure A; the red curve corresponds to structure B. Both curves are shifted towards lower values of the Hamming distance, illustrating a certain degree of similarity of multiple sequences encoding for the same structure. However, the Hamming distance between the sequences encoding for A and sequences encoding for B (green curve) is much larger and is very close to that of purely random sequences. Therefore, in sequence space, we can identify two groups of sequences that are similar within each group and dissimilar across the groups. Thus, the genomes of our model organisms can be classified according to their membership in the two well-defined groups of sequences, which is our model analogue of genome-based taxonomy. It is interesting to note that since our model is purely divergent and lateral gene transfer is not allowed, the evolving lines (or species) of organisms remain isolated, each evolving around its own DPS.

**Structural similarity network of evolved proteins.**

Now we turn to the discussion of structures of evolved proteins. An important global characteristic of the set of evolved proteins is the protein domain universe graph (PDUG) [2]. In this graph, non-homologous proteins are linked by an edge if their structural similarity score exceeds a certain threshold. It is known [2] that in natural proteins, the size of the largest cluster (giant component) of the PDUG abruptly shrinks at some value of the threshold, similar to the percolation transition. The degree distribution of the graph, i.e. the probability *p(k)* that a protein has *k* structurally similar neighbors, is a power law at the transition point. The scale-free character of this graph is believed to be a consequence of divergent evolution [2,30,31] as suggested by simple phenomenological "duplication and divergence" models [2]. Therefore, it is important to test whether our model can reproduce the global features of the natural protein universe that are manifest in the unusual properties of the PDUG.

Here we plot the PDUG of evolved proteins using Q-score – the number of common contacts between a pair of proteins – as a structural similarity measure [31]. The degree distribution of the evolved PDUG at similarity threshold *Q*=17 (the mid-transition in giant component of the evolved graph, see Supplementary Figure 4) is shown in Figure 7. The degree distribution plot clearly shows that the graph consists of two components, a scale-free-like component at lower *k*, and a small but very highly connected component at high *k*. As a control, we computed *p(k)* for a divergent model without the genotype-phenotype feedback, with the fixed death rate of organisms equal to the death rate in the exponential growth regime of evolution model. The degree distribution of the PDUG obtained in this control simulation where death rate is constant and independent of the stability of evolving proteins is shown in Supplementary Figure 5. The control graph is weakly connected, indicating randomness of the discovered structures. The degree distribution of the control graph is well approximated by a Gaussian distribution, in contrast to the one obtained from evolution simulation or the real PDUG [2].

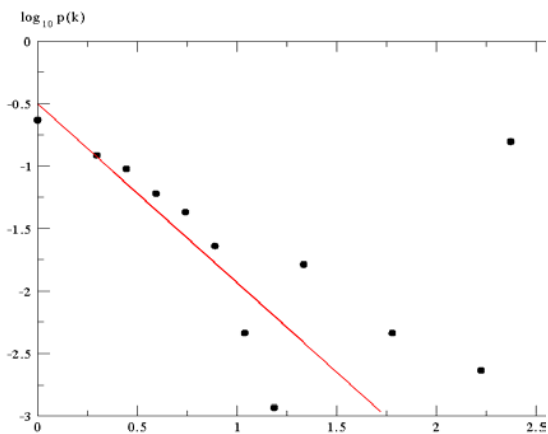

**Figure 7** *Degree distribution of structure similarity graph (PDUG) of the surviving populations in the evolution model. The similarity threshold was set to Q=17 corresponding to the transition point in the largest cluster size (the giant component) of the graph. The slope of the linear approximation is -1.4 for log k < 1.75.*

Therefore, evolutionary selection has a profound effect on the global structure of evolved protein universe. In the model, the structural similarity graph (PDUG) splits into a scale-free-like part and a highly connected part, corresponding to the DPS, populated by many dissimilar sequences.

Our simulations predict that new folds emerge as offsprings of DPS, and in this picture the DPS serve as prototypes of the first ancient folds. Following this logic one should expect that ancient protein folds, being closer to prototypical DPS, should be highly clustered and more connected than later diverged folds. To test this prediction we analyzed the subgraph of the PDUG corresponding to last universal common ancestor (LUCA) domains [32]. There are 915 LUCA domains. We compared the connectivity and clustering coefficient in the PDUG subgraph corresponding to LUCA domains with distributions for the same characteristics for 915 randomly selected domains as a control. The null hypothesis is that a random subset of protein domains has connectivity and clustering coefficients similar to that of the LUCA domains. In Supplementary Figure 6 we present the histograms of mean connectivity $<k>$ (average degree of the node) and clustering coefficient $C$ found in 20000 subsets of N=915 randomly chosen protein domains (out of total of 3300 DALI domains constituting the PDUG, see [2]). For random subsets of 915 domains from the PDUG, $<k>$=2.91, $<C>$=0.197 while the average values of the same parameters for the 915 LUCA protein domains: $<k>$=4.61, $<C>$=0.267. The values of $<k>$ and $<C>$ for the LUCA domains are statistically much greater than corresponding values for the random subsets (see Supplementary Figure 6), yielding extremely low $p$-values ($p<10^{-10}$) that LUCA domains are connected and clustered just as a random subset of the PDUG (assuming Gaussian distributions of mean connectivities and clustering coefficients for random subsets of the PDUG in Supplementary Figure 6). This proves that LUCA domains are statistically more connected and clustered than an equivalent set of random protein domains as predicted from our simulations.

In Figure 8, we summarize the divergent evolution scenario as observed in our model. Divergence and selection lead to the infrequent discovery of new protein folds (dashed circles). Within these folds, mutations result in the formation of protein

(super)families. The size of protein families steadily increases with time, so older families are generally larger. However, fold formation can occur at any time, branching off any family, so the newly formed families will be necessarily small. At the same time, the structures corresponding to superfamilies are all pairwise similar to each other and for that reason they are highly clustered in the PDUG. Therefore, at any moment, the snapshot of the evolving protein universe will comprise tightly clustered families of all sizes. This picture of protein families that are ''tightly knit'' within each fold leads to a prediction of a peculiar property of the PDUG: that each node (i.e. protein domain) with connectivity k is primarily connected with nodes of similar connectivity – members of its own fold family.

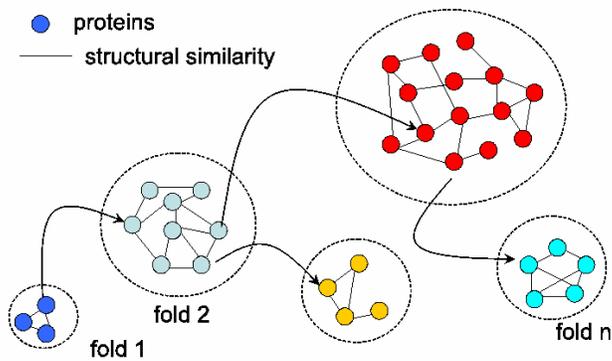

**Figure 8.** *Schematic representation of the formation of protein folds and superfamilies by punctuated jumps in the divergent model. Invention of new folds and their spread in population is a rare event whose time scale exceeds lifetime of an organisms and mutation time scale. On a shorter timescale mutations that do not change protein structure significantly occur and fix in the population. That gives rise to protein families (on the shortest time scales) or superfamilies (on time scales longer than mutational but shorter than fold innovation). Evolutionary time increases from left to right.*

To test this prediction we follow the approach proposed by Maslov and Sneppen [33]. Connectivity correlation in the PDUG is defined as the probability $P(k_1,k_2)$ that two evolved proteins that have $k_1$ and $k_2$ structural neighbors are structurally similar to each other, i.e. are themselves connected in the PDUG. To normalize $P(k_1,k_2)$, we created 1000 realizations of the rewired graph where each node has exactly the same connectivity as in the original graph of evolved structures, but with randomly reshuffled links to other nodes. The rewired graphs allow us to calculate the average value $P_r(k_1,k_2)$ and the standard deviation $\sigma_r(k_1,k_2)$ of the probability that nodes with connectivities $k_1$ and $k_2$ are connected in a particular network. In Figure 9a, we present the Z-score for connectivity correlations $Z(k_1,k_2)=(P(k_1,k_2)-P_r(k_1,k_2))/\sigma_r(k_1,k_2)$ for the natural PDUG. It follows from this plot that in PDUG, nodes of the similar degree tend to be connected to each other: high values of $Z(k_1,k_2)$ (red color) are grouped along the diagonal $k_1=k_2$. While at low $k$ this property is simply a consequence of the transitivity of the measure of structural similarity (if structure A is similar to B, and B is similar to C, then C must be similar to A), it is highly non-trivial to observe this property for highly-connected nodes. The pattern of connectivity correlations where similarly connected nodes tend to be connected to each other is very different from the one found in protein-protein interaction, communication, and social networks, where low-connected nodes tend to be connected with highly-connected hubs, but not to each other [33]. As seen from Figure 9b, our simple evolution model perfectly reproduces this unusual pattern of connectivity correlations. The reason for such unusual property of connectivity correlation is in the punctuated character of fold discovery and evolution both in the model and in real PDUG.

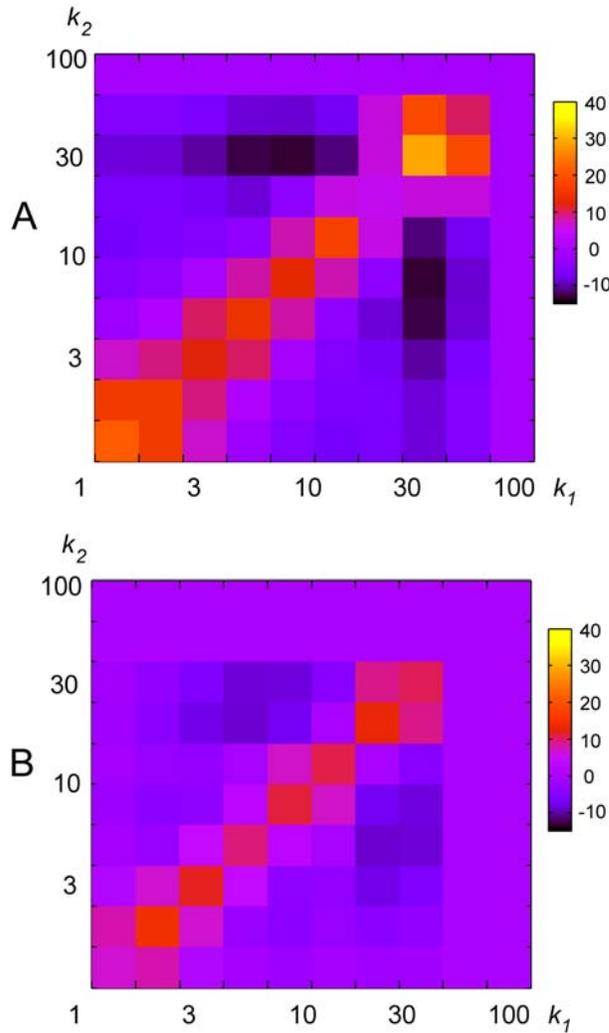

**Figure 9.** *Node degree correlations in evolved and natural PDUG. **(A)** Z-score for the probability $P(k_1,k_2)$ of the two nodes with degree $k_1$, $k_2$ being connected to each other in the natural protein domain universe graph. Unlike in other networks, nodes of similar degree tend to be connected. **(B)** The Z-score plot of $P(k_1,k_2)$ for the structure similarity graph obtained in the evolution model is remarkably similar to the actual one.*

**Discussion and Conclusions.**

In this work we introduced a microscopic physics-based model of early biological evolution which directly relates evolving protein sequences and structures to the life expectancy of the organism. We used a simple physical model of protein thermodynamics, and a simple Malthusian model of the population dynamics. The main assumption of our minimalistic model is that the necessary condition of survival of a living organism is that its proteins adopt their native conformations. Therefore, death rate of the organisms decreases when their proteins become more stable against thermal denaturation or unfolding. In other words we assume that *all genes of our model organism are essential*. Biological function is not explicitly present in the model, but protein stability is the necessary condition for its evolution. Genes in our model have high mutation rates, conducive to rapid innovation. As such our model can be directly applicable to (and can be experimentally tested on) the evolution of RNA viruses, which often encode for a handful of proteins, all of which are essential for the virus. The absence of error correction mechanism results in very high mutation rates and heterogeneous, quasispecies-like, populations of RNA viruses [34] [35] similar to what is found in this model. Rapid evolution makes RNA viruses an ideal system for experimental studies along the lines or our model, where the simulation algorithm propagates model organisms almost like an infected host cell produces new viral particles. The low number of genes (3-10 proteins per genome depending on simulations conditions, see Fig.5) observed in our model is in part related to the extremely high mutation rates, about 6 mutations per genome per replication. In modern life, such a high rate is observed only in populations of RNA viruses which lack the error correction mechanism. Remarkably, the genomes of RNA viruses are rather short and normally encode for less than ten proteins. More complex, DNA-based viruses and all cellular organisms invariably possess much lower mutation rates due to error correction and, correspondingly, longer genomes [36], in qualitative agreement with our model. Thus, our model suggests that protein stability requirements, together with mutation rates, play a crucial role in determining the size of the genomes of surviving organisms.

There is a common belief that the experimentally observed moderate stability of natural proteins is a result of positive selection for function. However no experimental proof for this conjecture is available. Rather, a circular argument that natural proteins are not extremely stable is offered to support this claim [37]. On the contrary, a recent study demonstrated that higher stability of a protein confers selective advantage to the protein by making it more evolvable, by enhancing its ability to tolerate more mutations and as a result evolve a new function [38]. A more plausible explanation of moderate stability of natural proteins is that it is a direct result of a tradeoff between stability in the native conformation and entropy in sequence space that opposes an evolutionary optimization beyond necessary levels [39]. We observe exactly this phenomenon in our model: while organisms with more stable proteins have selective advantage, the opposing factor – enormity of search in sequence/structure space – results in a compromise level of stability which corresponds to stable but not overstabilized proteins (see Figure 2c). By ''not overstabilized'' we mean here that for the same structure standard sequence design methods [40] [41] can provide sequences with $P_{nat}$ values that are much closer to 1 than observed in evolved model proteins (data not shown).

Unlike in many previous attempts, our model explicitly describes the interplay of evolution of individual genes and that of genomes (organisms) as a whole, since death of an organism leads to a complete loss of its genome. The model gives important insights into the interplay between molecular evolution, protein fold evolution, and population dynamics. In combination with selection pressure, random diffusion in sequence and structure spaces eventually leads to the discovery of specific structures, DPS, that are resistant to mutations and form very evolvable proteins. This, in turn, immediately leads to the "Big Bang" event whereby discovery of viable proteins is coupled to exponential population growth, as mutations are no longer a big threat to viability. The DPS persist over many generations, and may be infrequently replaced or augmented by other, even more favorable, structures, in a process similar to punctuated evolution. The remarkable separation of timescales between frequent mutations and rare DPS formation allows for the formation of the protein families and superfamilies. Our model suggests that the DPS may be superseded by more advantageous folds during evolution. A similar domain loss phenomenon has been discussed in [42] in the context of structure-based prokaryotic phylogenies.

The model and simulations presented here provide a quantitative first-principles description of evolution of the Universe of protein families. Despite simplicity of the structural model of proteins and phenotype-genotype relation invoked, it is able to quantitatively reproduce the power-law distributions that are observed in the natural Protein Universe. Earlier phenomenological models reproduced some aspects of power-law behavior, always at the expense of invoking dramatic assumptions about the dependence of the rates of gene duplication on the sizes of already existing gene families. Here no such assumptions are made as the model is fully microscopic in nature. Furthermore, our simulations are capable to reproduce not only ''power-law''-like behavior but also marked deviation from it. Indeed, as seen on Figure 4b there is an inflection point in the distribution of family sizes (blue curve) where the apparent slope changes. A similar inflection point was found in a recent clustering analysis of more than 7 million of Global Ocean Sampling sequences [43]. Strikingly the distribution of family sizes of evolved proteins (Figure 4a) features a noticeable inflection as well. It is not clear whether phenomenological duplication-growth models are capable to reproduce such fine details of the family size distribution.

The most intriguing (and relevant) question is the origin of the universally observed power-law distributions in our model. Clearly an explanation proposed in many phenomenological models [28] [3] is not applicable here because the rates of all processes, including gene duplication, are constant in the model and do not depend on sizes of already existing gene families. Therefore there are no ad hoc assumptions about the gene birth/death dynamics in the model that could result in power-law distributions. The only plausible reason may be that the underlying dynamics in sequence and structure spaces, coupled with selective pressure, is responsible for the emerging power law distributions. Indeed, our key finding concerns *dynamics* of fold discovery and death - that the lifetimes of DPS are power-law distributed (Figure 3b). The size of a protein family (and superfamily, on longer time scales) is proportional to DPS lifetime. Indeed, power law exponents for family size distribution and DPS lifetimes are very similar. While these observations are suggestive, a more detailed future analysis of our model will make it possible to find a definite answer as to the origin of ubiquitous power law distributions in sequence and fold statistics.

Several earlier studies modeled evolution of proteins by applying pressure directly on the proteins, assuming that the probability of replication of a protein in population of proteins depends on its molecular properties such as e.g. stability [10,11,22,39] or folding kinetics [19,44] or both [16]. In contrast, in the present model biological (or as will be argued below ''physiological'') constraints are applied to organisms as a whole, not to individual proteins. Evolutionary simulations and simple theory presented here highlight the importance of this distinction: the genome sizes are closely connected with maximum and average stability of evolved proteins. Therefore, biological pressure is ''distributed'' in the genome and all genes act in concert in response to it. Furthermore, no Dominant Protein Structures were found in earlier simulations [10], despite the fact that overall population of evolved lattice proteins was somewhat skewed towards more designable structures. In contrast, our key finding is that evolution of population is strongly coupled with protein evolution, as population growth is contingent upon discovery of a very limited set of protein structures. The difference here may be due to the fact that simpler, 2-dimensional lattice models were used in previous simulations, or due to the differences in how biological pressure is applied – on whole organisms here and on individual proteins in earlier works [10,45]

The presented model is markedly different from standard models of population genetics (PG) such as Fisher-Wright and quasispecies (QS)[46,47]. PG and QS models are phenomenological descriptions of evolution, attributing certain values of fitness for the genomes with predetermined combinations of alleles. These models conveniently sidestep the important question of the molecular origins of the change of fitness upon recombination or mutation. The genotype-phenotype relationship in our model is not phenomenological but physiological: when a gene product loses stability (and, by implication, functionality) the whole organism is likely to die. This assumption is justified by recent high-throughput experiments that use RNAi to determine the impact of gene knockout on phenotype [48,49]. As in an experiment where knockout of essential genes results in death of an organism, in our model deterioration of stability of any gene of an organism confers the lethal phenotype. In the present implementation our model assumes that all genes are essential. However this assumption can be relaxed (work in progress) making it possible to study differentially the impact of biological constraints on evolution of genes [50].

Another critical distinction between our approach and traditional phenomenological models is that in PG and QS approaches a single genotype *is assumed* to be advantageous[46,51]. While the outcome may be that genomes of the populations are peaked around the most fit one (as in the standard QS model [47]) or a broader distribution among genotypes may emerge (as in the ''survival of the flattest'' scenario [52]) it is always an implication of the key assumption that a certain genotype confers the highest fitness. In contrast, the present model makes no *a priori* assumptions about the fitness advantage of a certain genotype. Strikingly, sets of organisms distributed around dominant genomes and proteomes - species - emerge here *as a result of evolution* at longer evolutionary times. A key factor determining the emergence of species in this model is that productive evolution occurs only when the structural diversity of proteins collapses into a small set of Dominant Protein Structures.

Our model of natural selection is minimalistic and is limited in its scope. It does not take into account such important biological processes as horizontal gene transfer, gene recombination, sexual reproduction, ''death of a gene'' (via pseudogenisation) and Darwinian selection due to competition of populations for limited resources. Also, in order to make the minimum possible number of assumptions, the modern amino acid alphabet is used in the model, although it has been suggested that the amino acid alphabet itself had evolved over time [53] [54,55]. However, we believe that our model is an important step towards the unification of microscopic physics-based models of protein structure and function and the macroscopic (so far, phenomenological) description of the evolutionary pressure. Its extensions are straightforward and may include a more explicit consideration of protein function, protein-protein interactions and fitness function that rewards functional (and therefore, structural) innovations. Furthermore, since habitat temperature enters the model explicitly it can be used to study thermal adaptation of organisms as well as adaptation to variable mutation rates. This work is in progress.

**Model and Methods.**

**Population dynamics and genotype-phenotype relationships.** In our model, an organism is completely described by the set of its genes. The genetic code then defines amino acid sequences, and exact nature of the lattice protein folding model makes it possible to find the native structures of the encoded proteins. We assume that for an organism to function properly, it is imperative that its proteins spend a significant part of the time in their native conformations at a given environmental temperature. Let $P_{nat}^{(i)}$ be the thermodynamic probability that protein $i$ is in its native conformation (see Protein Model below). As a simplest approximation, we assume that the probability that an organism is alive is proportional to the lowest $P_{nat}^{(i)}$ across all of its proteins:

$$P_{alive} \propto \min_i P_{nat}^{(i)} \quad , \tag{M1}$$

i.e. longevity of an organism is determined by the least stable protein in the genome ("weakest link" model).

Our model of population and genome dynamics includes four elementary events: 1) random mutation of a nucleotide in a randomly selected gene, with constant rate $m$ per unit time per DNA length; mutations leading to the stop codon are rejected to ensure the constant length of protein sequences; 2) duplication of a randomly selected gene within an organism's genome, with constant rate $u$; 3) birth of an organism via duplication of an already existing organism with constant rate $b$ (the genome is copied exactly); 4) death of an organism, with the rate $d$ per unit time (Figure 1). For simplicity, we do not allow for the formation of pseudogenes or any other mechanism of removal of the genes from a genome; in every organism the number of genes increases (or remains constant) with time. However, the average number of genes per organism in the population can either decrease or increase due to enhanced survival of organisms with shorter (longer) genomes.

Condition (M1) translates into the dependence of organism death rate $d$ on the stability of its proteins:

$$d = d_0 \left(1 - \min_i P_{nat}^{(i)}\right), \tag{M2}$$

where $d_0$ is the reference death rate. This relation gives rise to an effective selection pressure on proteins since organisms which have at least one unstable protein live shorter and thus produce less progeny. This simple, direct and physically plausible relationship between the genotype (thermodynamic properties of the proteins) and the phenotype (life expectancy) is the key novel feature of our model. Another implication of this relationship is in the ''collective punishment '' effect that genes do not evolve independently: a very unfavorable mutation in a gene will likely lead to a quick death of an organism, so its complete genome will not be able to proliferate. Such cooperativity creates an important selection pressure towards mutation-resistant genes encoding stable and evolvable (see below) proteins. Interestingly, purely physical factors ensure that resistance to mutations, evolvability of a new function and thermostability are well correlated [24,38], so little or no trade-off may be needed to satisfy both requirements. To ensure that a sufficient selection pressure is applied, we set $d_0=b/(1-P_{nat}^{(0)})$, where $P_{nat}^{(0)}$ is the native state probability of a protein encoded by the primordial gene, which is the single gene in all organisms from which evolution runs start. Therefore, the Malthus parameter $b-d$ of population growth is zero for neutral mutations (not changing $P_{nat}$ with respect to the primordial sequence), positive for favorable mutations which increase $P_{nat}$, and negative for deleterious mutations. In principle, the relationship between growth rate and protein stability can be experimentally verified by analyzing the growth rate of bacteria at elevated temperatures. While the exact biochemical mechanisms leading to slower replication and eventual death are complicated, they all originate in the loss of protein function or enzymatic activity due to thermal denaturation [56]. A sequence evolution model, also using the protein stability $P_{nat}$ as fitness parameter has been recently proposed by Goldstein and coworkers [54].

**Simulation algorithm**

In our model, each organism is represented by a list of its genes, 81-nucleotide sequences that are translated into amino acid sequences according to the genetic code. There can be up to 100 genes per organism; the gene duplication rate is chosen so that this limit is never reached in a simulation; typically, organisms have less than 10 genes each at the end of a simulation. Initially, 100 organisms are seeded with one and the same primordial gene; $P_{nat}^{(0)}$ is the native state probability of the protein encoded by the primordial gene.

At each time step of the evolution, each organism can undergo one of the five events: no event at all, or the four events described in the main text (duplication of an organism with probability $b=0.15$, death with rate $d$, gene duplication with probability $u=0.03$, point mutation of a randomly chosen gene with probability $m=0.3$ per gene). The organism death rate is calculated according to eq. (M2), $d = d_0 \left(1 - \min_i P_{nat}^{(i)}\right)$, with $d_0=b/(1-P_{nat}^{(0)})$.

Every 25 time steps, an entire set of genes of all currently living organisms is recorded for analysis. The simulation stops after 3000 time steps. Whenever the population size $N$ exceeds 5000, we randomly remove $N$-5000 organisms to ensure constant population size, simulating a turbidostat; despite the artificially constrained population size, the growth regime remains exponential.

**Protein Model.**

To simulate the thermodynamic behavior of evolving proteins, we use the standard lattice model of proteins which are compact 27-unit polymers on a 3x3x3 lattice [57]. The residues interact with each other via the Miyazawa-Jernigan pairwise contact potential [58]. It is possible to calculate the energy of a sequence in each of the 103346 compact conformations allowed by the 3x3x3 lattice, and the Boltzmann probability of being in the lowest energy - native - conformation,

$$P_{nat}(T) = \frac{e^{-E_0/T}}{\sum_{i=0}^{103345} e^{-E_i/T}}, \qquad (M3)$$

where $E_0$ is the lowest energy among the 103346 conformations, $E_i$ are the energies of the sequence in the remaining 103345 conformations and $T$ is the environmental temperature (in the simulation, we assumed $T=0.5$ in Miyazawa-Jernigan dimensionless energy units).

**Analytic calculation of genome sizes of model organisms**

Suppose each genome has $N$ genes, and the fitness of the entire genome is then defined by $f = \min\{P_{nat}^{(1)}, \ldots, P_{nat}^{(N)}\}$. Based on the sequence design simulation, we find that it is a reasonably good approximation to assume that in our lattice model the distribution of stability $P_{nat}$ of a lattice protein after a point mutation (i) does not depend on the stability before the mutation, and (ii) is uniformly distributed between 0 and 1. Performing a point mutation, we can either mutate the gene with the lowest fitness value, with probability $\frac{1}{N}$ (case A), or select any one of the other, more stable, genes and mutate it with probability $(1-\frac{1}{N})$, case B.

In case A, because the mutated gene was the original least-stable gene, there are two possible outcomes after the mutation: (i) If the new gene fitness value is less than $f$, then this new gene fitness value would be the new minimum among all gene fitnesses in the genome, therefore this new fitness will become the fitness of the new genome. This occurs with a probability $f$, and since the new fitness follow a uniform distribution in the region [0, 1], the expectation value in this case is $f/2$. So this part's contribution to the expectation value of the new genome fitness is $A_1 = \frac{1}{N} f \frac{f}{2}$. (ii) If the new gene stability is greater than $f$, which happens with a probability of $(1-f)$, we can calculate the probability distribution for the new genome fitness being $x$ is

$$p(x) = \left(\frac{1-x}{1-f}\right)^{N-1} \text{ (for } f<x<1\text{)}.$$

This equation means when one gene has fitness $f<x<1$ and is the new minimum, also under the condition that all fitness are within the region of [f, 1], the probability for all the other (N-1) genes has to have fitness greater than x is $\left(\frac{1-x}{1-f}\right)^{N-1}$. The multiplicity of this condition is $M_{N,f} = \frac{N}{1-f}$. This is because that we can pick any one of the N genes to be the new least-fit-gene, and the fitness value is within the region [f, 1] with uniform probability distribution. The contribution for the new genome fitness in this situation is therefore the total product of the probability of this situation $\frac{1}{N}(1-f)$, the multiplicity $M_{N,f}$, and the expectation value $\int_f^1 xp(x)dx$

$$A_2 = \frac{1}{N}(1-f) * M_{N,f} \int_f^1 xp(x)dx.$$

In the case B, we also have two possible situations, situation $B_1$ states when the mutated gene has a fitness less than $f$, and situation $B_2$ states when the mutated gene has a fitness greater than $f$. In situation $B_1$, similar to the derivation in case $A_1$, the probability for the new stability to be smaller than $f$ is $f$, and the expectation value of the new genome fitness in this case is $f/2$. So we have $B_1 = \left(1 - \frac{1}{N}\right) f \frac{f}{2}$.

In situation $B_2$, if the stability of the mutated gene is greater than $f$, then the original gene with stability $f$ would remain the least stable in the genome. Therefore the genome fitness in this situation is still $f$, so the value of $B_2$ reads $B_2 = \left(1 - \frac{1}{N}\right)(1-f)f$, where $\left(1 - \frac{1}{N}\right)$ is the probability to choose one of the (N-1) genes with stability greater than $f$, *(1-f)* is the probability to mutate this gene with fitness greater than $f$, and $f$ is the expectation of the final fitness under this condition..

Finally, summing up $A_1, A_2, B_1, B_2$, we obtain the expectation value of the genome fitness after one point mutation:

$$<f'> = \frac{2 - 4f + f^2(2 + 3N - N^2)}{2N(N+1)} + f$$

Now, if the average genome fitness after a single point mutation must be greater than the original fitness, the condition <f'>-f > 0 must be satisfied. Solving this inequality, we find an upper limit on the number of genes in a genome, eq. (2).

**Family and Superfamily Size Estimate for Real Proteins**

We take sequences of all structurally characterized domains from HSSP[59]. We use BLAST[60] with threshold $10^{-10}$ to identify all sequences with significant homology to each HSSP domain in a non-redundant sequence database NRDB90[61]. We combine each set of sequences with homology into a single gene family. The number of non-redundant sequences matching the domain is the number considered in that family. We then use cross-indexing between NRDB90[61], Swiss-Prot[62] and InterPro[63] to define the set of different functions each gene family performs.The number of different functions as defined by InterPro becomes the number of superfamilies folding into the same domain.

**Family and Superfamily Size Estimate for Model Proteins**

In the model, the superfamily size is defined as the number of nonhomologous sequences with all mutual pairwise Hamming distance of 16 or more (i.e. 40% sequence identity or less) having the same native conformation. The family size is defined as the number of all sequences folding into a given structure, without removing the homologous sequences.

**Protein domain universe graph**

To construct the protein domain universe graph (PDUG) from the simulation data, we consider only the nonhomologous amino acid sequences. The selection is based on the Hamming distance between the sequences, which should exceed 18 (i.e., less than 33% sequence identity).

To calculate the structure similarity in the PDUG, we use the Q-score similarity measure. The Q-score measure between the two structures $i$ and $j$ is the number of all pairs of monomers $(k,m)$ that are in contact both in structure $i$ and structure $j$. As there are always 28 contacts in compact 27-mers, Q-score varies from 0 for completely dissimilar structures to 28 for two identical structures. The Q-score is analogous to the DALI Z-score, used as a structural similarity measure for real proteins.

**Definition of LUCA domains**

The simplest construction of the LUCA that still yields useful information is the delineation of the very old domains. Any domain shared by the three kingdoms of life can be placed in the last universal common ancestor (LUCA)[64]. If any such domain were not placed in the LUCA, multiple independent discovery (or horizontal transfer) events would be required to explain the occurrence of this domain in all kingdoms. The "extra" evolution involved in this case would result in a less parsimonious scenario. Inclusion of other domains is more probabilistic and depends on the exact form and method of parsimony construction used.[64] We thus define the structural content of the LUCA to be all domains that have homologs in at least one archaeal, at least one prokaryotic and at least one eukaryotic species. This yields approximately a third of the PDUG members.

**Acknowledgments**

The authors acknowledge stimulating discussions with I.N. Berezovsky and financial support from the NIH.

**Supporting information**

**Supplementary Figure 1.** Mean genome size (number of genes per organism) in an exponentially growing population is almost constant (red curve) due to the balance between gene duplication and selection pressure. In a simulation run where population becomes extinct, the genome size grows linearly with time (blue curve).

**Supplementary Figure 2. (a, top panel).** Abundance of different structures in the proteomes in an unsuccessful evolution run as a function of time. Red corresponds to abundant structures, and cyan to rare or nonexistent ones. **(b, middle panel)** Size of population as a function of time. **(c, bottom panel)** Mean native state probability $<P_{nat}>$ as a function of time. Dominant protein structures are never found in this run, resulting in extinction of the population due to random diffusion in sequence space.

**Supplementary Figure 3.** Probability distributions of lattice protein stability $P_{nat}$ after a point mutation in sets of sequences with a given stability $<P^0_{nat}>$. The peak at $P_{nat}=<P^0_{nat}>$ corresponds to mutations that do not alter the stability; at high $<P^0_{nat}>$, the long constant-level tail of the distribution makes it possible to approximate the distribution by a uniform one. For each plot, we performed all possible 19*27=513 mutations in 100 different sequences with $| P^0_{nat} - <P^0_{nat}> | < 0.03$.

**Supplementary Figure 4.** Fraction of the giant component of the PDUG as a function of similarity cutoff $Q$ for evolution simulations.

**Supplementary Figure 5.** Degree distribution of structure similarity graph (PDUG) in a control simulation where genotype-phenotype relationship does not exist. The similarity threshold was set to $Q=17$ corresponding to the transition point in the largest cluster size (the giant component) of the graph.

**Supplementary Figure 6**. Probability distribution of the average connectivity (a) and clustering coefficient (b) for random subsets of 915 protein domains from the PDUG, and the value of these parameters $<k>=4.61$ and $<C>=0.267$ for the LUCA domains (red line). The distribution is drawn over 20,000 random selection of 915 subsets out of total 3300 PDUG domains.

**Supplementary Figure 7**. Average number of genes per organism at low mutation rate m=0.1 (red curve) and at a higher mutation rate m=0.2 (black curve) as a function of time in typical evolution runs. Organisms evolving at a higher mutation rate evolve shorter genomes. The temperature is $T=0.8$.

Supplementary Figure 1

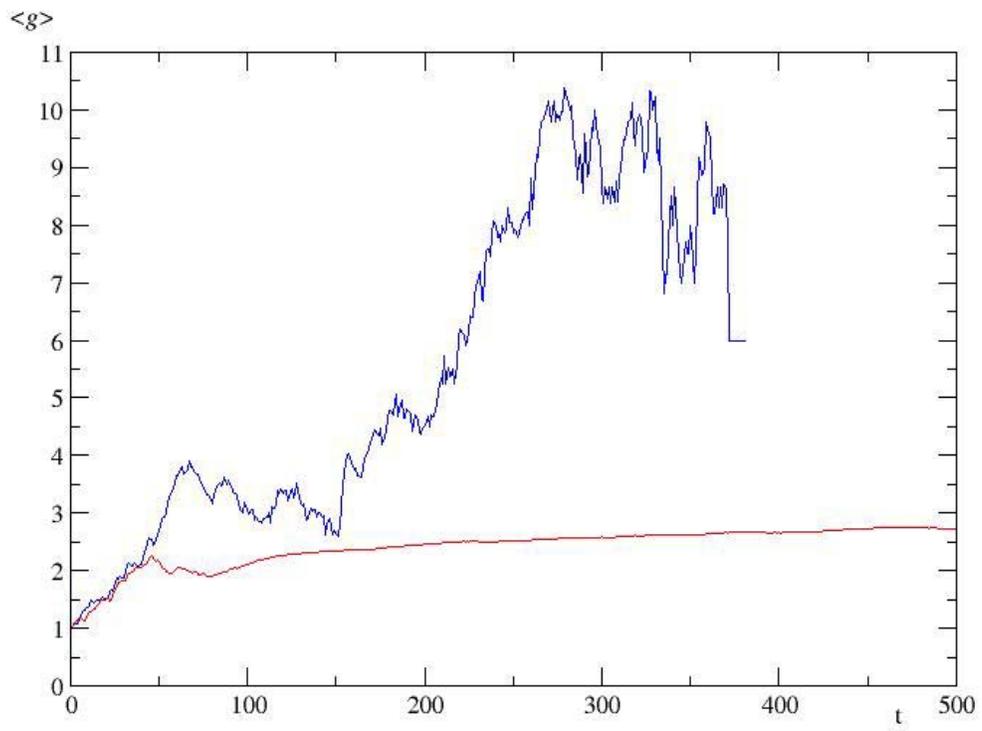

Supplementary Figure 2

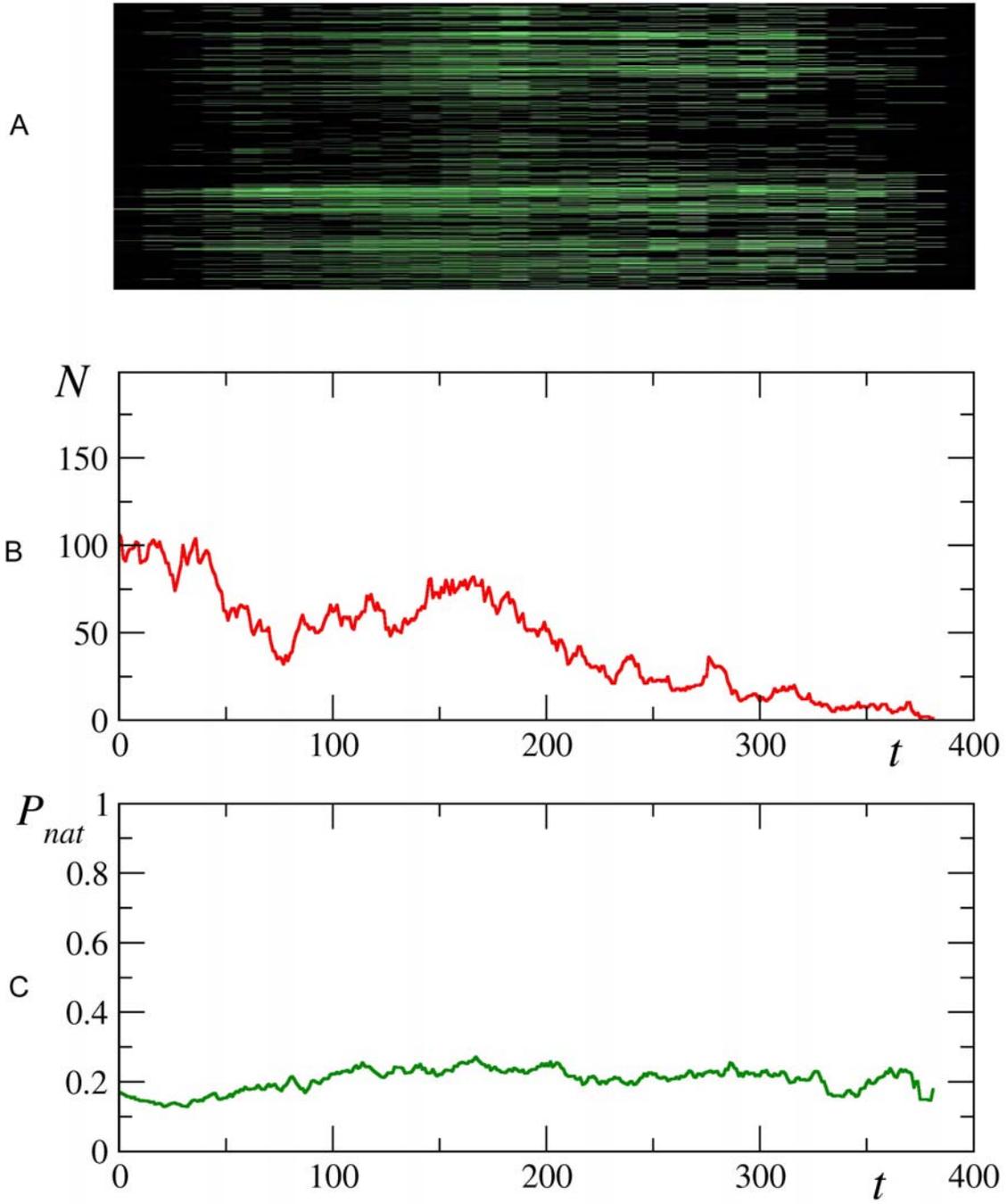

Supplementary Figure 3

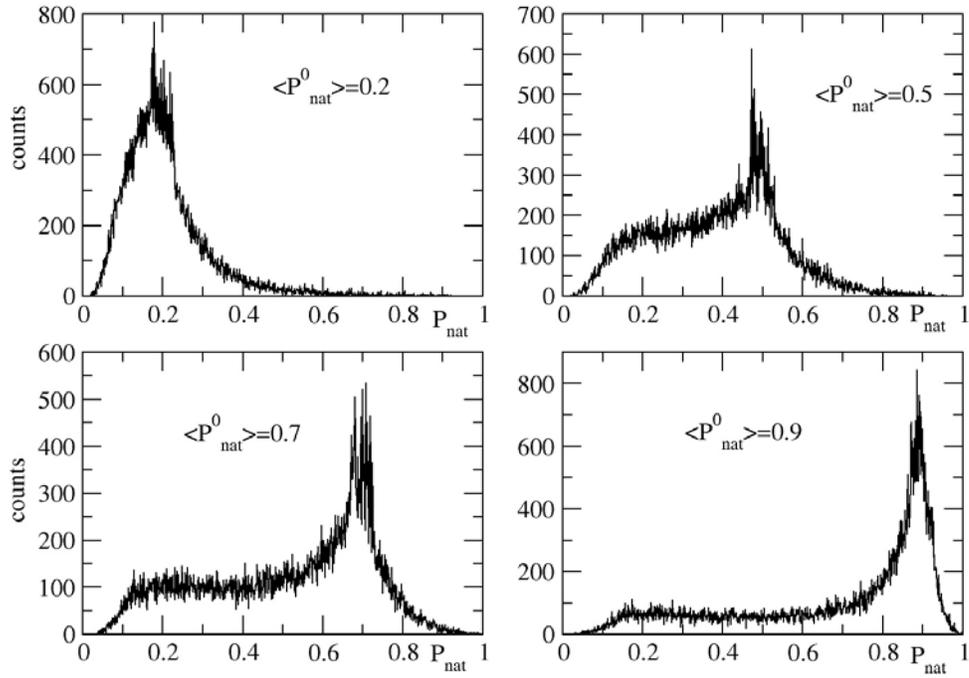

Supplementary Figure 4

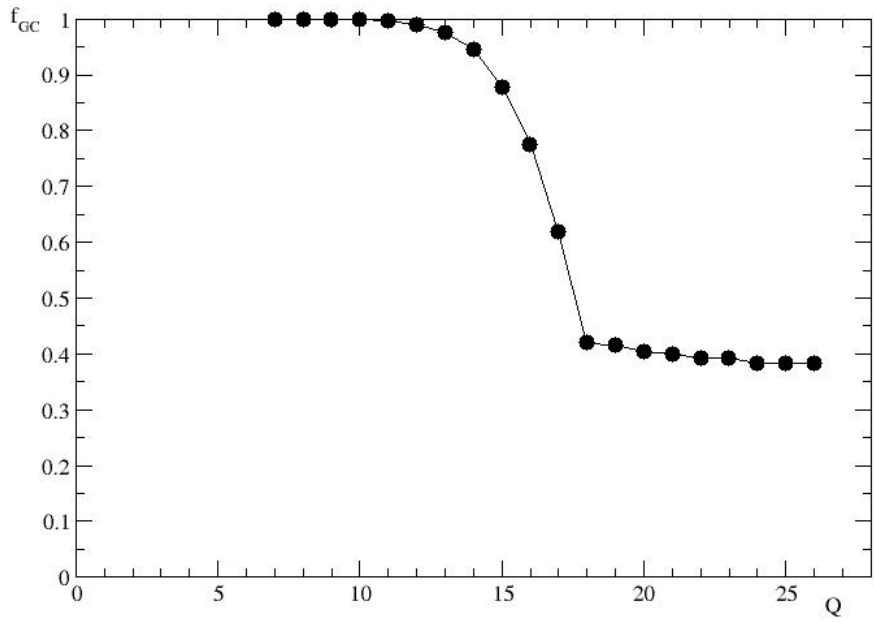

Supplementary Figure 5

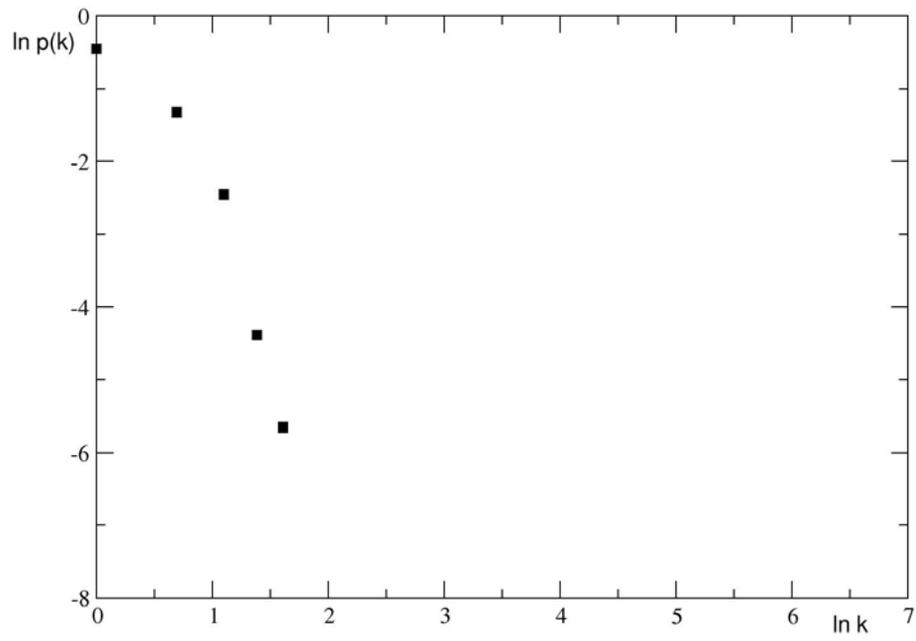

Supplementary Figure 6a

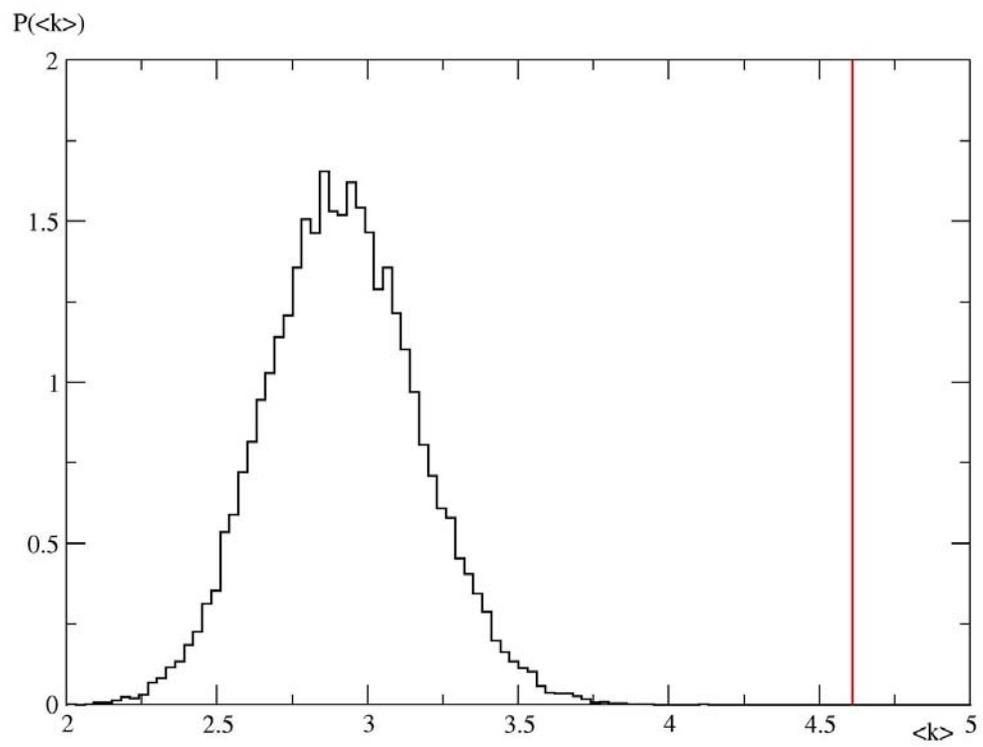

Supplementary Figure 6b

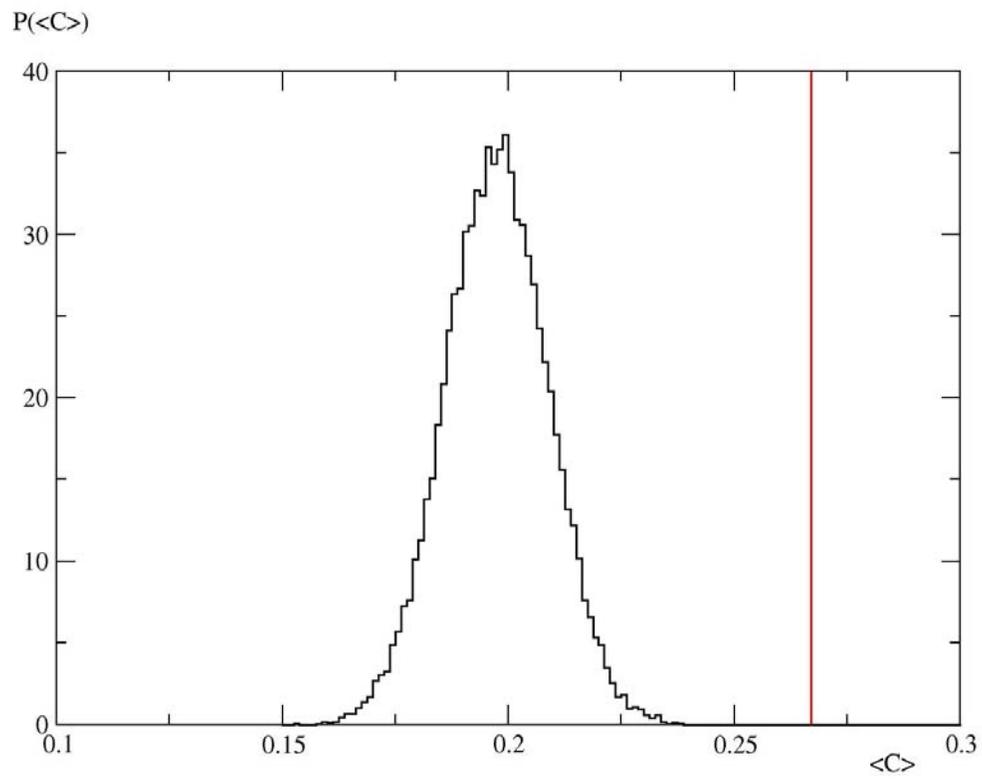

Supplementary Figure 7.

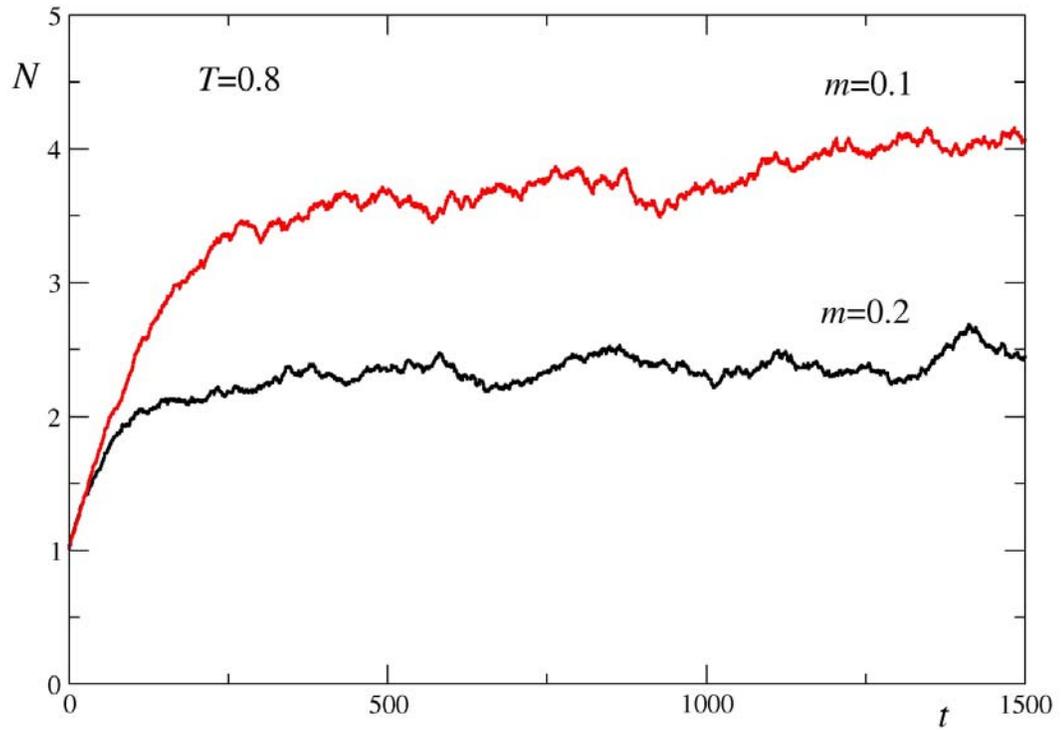